
\magnification=\magstep1
\baselineskip=18pt

\raggedbottom

\def\d{\partial}

\def\newadd{{\it current address}:  Program in Atmospheric and Oceanic
Sciences, Department of Astrophysics, Planetary, and Atmospheric
Sciences, Campus Box 311, University of Colorado, Boulder, CO 80309 }

\centerline{\bf Hamiltonian Maps and Transport in Structured
Fluids}

\medskip

\centerline{Jeffrey B. Weiss\footnote*{\newadd}}
\centerline{National Center for Atmospheric Research}
\centerline{P.O. Box 3000}
\centerline{Boulder, CO \ 80307}

\medskip

Structures such as waves, jets, and vortices have a dramatic impact on
the transport properties of a flow.  Passive tracer transport in
incompressible two-dimensional flows is described by Hamiltonian
dynamics, and, for idealized structures, the system is typically
integrable.  When such structures are perturbed, chaotic trajectories
can result which can significantly change the transport properties.
It is proposed that the transport due to the chaotic regions can be
efficiently calculated using Hamiltonian mappings designed
specifically for the structure of interest.  As an example a new map
is constructed, appropriate for studying transport by propagating
isolated vortices.  It is found that a perturbed vortex will trap
fluid parcels for varying lengths of time, and that the distribution
of such trapping times has slopes which are independent of the
amplitudes of both the vortex and the perturbation.

\medskip

\centerline{August 1993}

\medskip

\centerline{to be published in Physica D}
\centerline{Proceedings of the NATO and EGS Workshop on}
\centerline{\it Chaotic Advection, Tracer Dynamics, and Turbulent
Dispersion}
\centerline{Sereno Di Gavi, Italy, May 1993.}

\vfil \eject

\noindent{\bf I. Introduction}

Many fluid flows are highly structured, containing a variety of
vortices, waves, jets, and fronts.  These structures can exist over
an extremely wide range of parameter values, from nonturbulent flows
with relatively weak forcing such as Rayleigh-Benard convection beyond the
first bifurcation, to extremely turbulent flows such as planetary
atmospheres and oceans.  Structures can have a profound impact
on the transport properties of flows through a variety of mechanisms;
jets can act as barriers to transport, while vortices and waves can
trap fluid parcels and carry them large distances.   These phenomena
are of great importance in geophysical flows.  The Gulf Stream can act
as a barrier to transport, affecting both heat and salinity transport
in the North Atlantic [1], the stratospheric polar night
vortex traps fluid within its boundaries where chemical reactions
result in ozone depletion [2], and vortices in the ocean can
carry water far from its original source affecting the overall
transport in the ocean [3, 4].

Studying the transport properties of a fluid requires the use of two
different descriptions of a flow.  The Eulerian description specifies
the fluid properties (velocity, temperature, etc.) in a fixed
reference frame, while the Lagrangian description specifies properties
in a frame moving with each fluid parcel.  The Eulerian velocity is
$\vec u_E(\vec x,t)$ where $\vec x$ can be, for example, a position
fixed in the frame of the laboratory for a laboratory experiment, or a
position fixed to the Earth for a geophysical flow.  Lagrangian
properties are determined by $\vec X(\vec x_0,t)$, the position at
time $t$ of the fluid parcel labeled by its initial position $\vec
x_0$, $\vec X(\vec x_0,0) \equiv \vec x_0$.  The relationship between
the two descriptions is the fact that an ideal fluid parcel is
transported by the local Eulerian velocity, $$ d \vec X/ dt = \vec
u_E(\vec X, t). \eqno(1)$$

In this paper we shall only be concerned with the transport of ideal
passive tracers that are described by (1).  Real
contaminants may have other forces acting on them such as drag or
buoyancy which modify the relation (1).  These effects are
addressed by other contributions in this volume.  Another
simplification we shall make is that we only consider two-dimensional
flows.  Many geophysical flows are constrained by rapid rotation and
stable stratification, resulting in flows which are approximately
two-dimensional.  In addition, some laboratory flows have a symmetry
which makes them effectively two-dimensional.  The question of
extension to three dimensions is certainly interesting, but we defer
it to later work.  In section II of this paper we discuss the general
notion of using Hamiltonian maps to study transport.  In section III
we introduce a new map to study the geophysically important phenomenon
of transport by isolated vortices.

\bigskip

\noindent{\bf II.  Hamiltonian Maps}

The typical pathway for theoretically studying transport is to
determine $\vec u_E$ and then use (1) to study the
Lagrangian behavior of fluid parcels.  In many situations, accurately
determining the Eulerian flow field is a difficult task in its own
right.  Laboratory measurements and geophysical observations are often
too sparse to determine $\vec u_E$ over a large region.  For simple
flow situations one may be able to determine the Eulerian flow
analytically through such tools as bifurcation theory.  In the case of
complex flows, however, one must usually resort to numerical
simulation.

Even if one knows the Eulerian flow, studying transport requires
integrating (1) for a long time using a large number of
initial conditions.  These computations, even for a single set of flow
parameters, can often be too expensive to be currently practical.
In this paper we discuss the use of Hamiltonian maps to efficiently
calculate transport properties of two-dimensional structured flows.
One such map has been previously used to study transport in waves
[5, 6]. Here we present the viewpoint that this technique is
applicable to a wider variety of structured flows, with each class of
structures requiring a different specific Hamiltonian map.

The relationship between Hamiltonian dynamics and transport is due to
the incompressibility of the flow [7].  In incompressible
two-dimensional flows the Eulerian velocity $\vec u_E = u \hat x + v
\hat y$ is determined by a streamfunction $\psi(x,y,t)$:
$$    u = -\d \psi / \d y, \quad v = \d \psi / \d x.
\eqno(2)
$$
Equations (1) and (2) imply that the parcel
trajectories $\vec X = x \hat x + y \hat y$ are given by Hamilton's
equations with $\psi$ acting as the Hamiltonian:
$$
\dot x =  -\d \psi / \d y, \quad \dot y = \d \psi / \d x.
\eqno(3)
$$

The entire theory of Hamiltonian dynamics can now be applied to
questions of transport in two-dimensional incompressible flows
[8].  One important fact is that if $\psi$ is independent
of time then (3) is integrable and the fluid trajectories
are regular.  Steadily propagating patterns are also integrable since
the streamfunction is stationary in a moving reference frame.  If an
integrable streamfunction has fixed points connected by a separatrix
(i.e.  a homoclinic or heteroclinic connection) then a periodic
perturbation will typically result in chaotic fluid trajectories.
Furthermore, the chaos will have the usual fractal structure of
resonant islands, KAM curves, and cantori.  An interesting aspect of
this is due to the fact that the phase space in (3) is
the physical space of the fluid.  Thus, the phase space structures
appear in the fluid itself and can be observed experimentally
[8].

It is often the case that a pure idealized fluid structure can be
described by a stationary or steadily propagating streamfunction.  The
transport due to this idealized structure is not too difficult to
discern, as the streamfunction is integrable and the trajectories are
regular.  If the amplitude of a structure becomes sufficiently large,
then the streamfunction can undergo a bifurcation creating fixed
points connected by separatrices, and the separatrices can divide the
fluid into regions which have very different transport behavior.  For
example, if the amplitude of a single frequency traveling wave is
sufficiently large, then the fluid is divided by a separatrix into two
regions: a region where parcels are trapped by the wave and carried
long distances, and a region where parcels flow backwards with respect
to the wave [9].  Two parcels starting nearby, but on
different sides of the separatrix will thus have very different fates.

An important point is that the qualitative nature of transport by
fluid structures is determined by the topology of the separatrices in
the time-independent idealized structure, i.e. how the separatrices
are connected to each other.  Geometrical features such as the exact
shape or location of a separatrix will only affect quantitative
aspects of the transport.  Thus, one has a certain freedom in
characterizing the streamfunction of a structure.  As long as a
streamfunction correctly captures the topology of separatrices it will
qualitatively reproduce the transport.  Additionally, if the
streamfunction's geometry can be varied by changing parameters then
one can study how quantitative aspects of transport depend on the
geometry.  Thus, one can profitably study transport using an
approximate streamfunction, bypassing the difficulty of accurately
determining the Eulerian flow.

Once the idealized structure is modeled by an integrable
streamfunction the next step is to study the effect of perturbations.
Here we use the hypothesis that the chaotic behavior of weakly
perturbed integrable Hamiltonian systems is relatively insensitive to
the details of the perturbation.  This hypothesis is extremely useful
in that it both allows one to choose particularly convenient
perturbations and it frees one from having to study in detail the
types of perturbations which affect a given structure in its natural
environment.  By carefully choosing a periodic perturbation one can
obtain a system which can be analytically integrated over a single
period of the perturbation, transforming the original set of
Hamiltonian ordinary differential equations into a Hamiltonian map.
This is advantageous since it is significantly easier and faster to
compute iterations of maps than integrations of ordinary differential
equations.  This procedure was originally used to study chaotic
behavior in the pendulum where it results in the standard map
[10, 11].

While the hypothesis of insensitivity to details of the perturbation
is certainly true for periodic perturbations, it is unclear how valid
it is for other types of perturbations.  Beigie, et.al. [12],
studied transport in quasiperiodically perturbed flows, and discussed
how their ideas apply to more general perturbations.  Babiano, et.al.
[13], studied advection in flows with more complex time
dependence, looking at both collections of point vortices, whose
motion is chaotic, and at two-dimensional turbulence.  In all these
cases, the separatrices of the unperturbed structures are found to
break producing chaotic behavior.  However, it is still unknown how
sensitive transport behavior in chaotic regions is to the
time-dependence of the perturbation.  The method we present here
allows one to efficiently study transport in a periodically perturbed
flow.  The use of this technique should help in understanding the
differences between periodic and non-periodic flows.

Thus, the strategy for efficiently calculating transport in the
chaotic regions of a structured flow is as follows.  First, one
constructs a simple time-independent streamfunction whose topology of
separatrices matches those of the structure of interest.  Second, one
adds a periodic perturbation which can be analytically integrated,
creating a Hamiltonian map.  The resulting map can then be iterated
numerically for very long times, using large numbers of initial
conditions, and for a wide range of parameter values.

\bigskip

\noindent{\bf III.  Isolated Vortices}

Isolated vortices appear in many regions of the Earth's ocean and play
a significant role in the ocean's transport of heat and salinity.  One
well known type of ocean vortex is a Gulf Stream ring, created when a
meander in the Gulf Stream grows in amplitude and pinches off
[3].  The pinched-off meander becomes a vortex whose interior
contains water that was originally on the opposite side of the jet.
Since the Gulf Stream acts as a barrier to heat and salinity
transport, these vortices provide a mechanism for transport across the
jet.  Gulf Stream rings typically have diameters of a few hundred
kilometers, and lifetimes of 1-4 years.  Similar vortices also detach
from intense currents in other parts of the ocean.  Another type of
vortex, called a Meddie (Mediterranean eddie), is found in the North
Atlantic and is associated with outflow from the Mediterranean Sea
[14].  Meddies have large heat and salinity anomalies,
diameters of about 100 km, and have been tracked for up to 2 years.
It is thus of great geophysical interest to learn how water is
exchanged between the interior and exterior of isolated vortices.

Isolated vortices on the Earth travel westward due to the
``beta-effect'', the variation in the Coriolis force with latitude
[15, 16].  In a frame comoving with the vortex, the
streamfunction of an idealized vortex has a single fixed point with a
homoclinic orbit [4] (Figure~1).  Fluid parcels inside
the homoclinic orbit are trapped and carried with the vortex while
parcels outside are left behind by the vortex.  Parcels outside the
vortex but on the stable manifold of the fixed point are carried with
the vortex forever, but these parcel form a set of measure zero and
are not significant.  What is significant is that parcels close to the
stable manifold slow down (in the comoving reference frame) on
approaching the fixed point and are thus carried with the vortex for a
while before being left behind.

Ocean vortices are subject to a variety of perturbations: they emit
Rossby waves as they propagate producing a periodic perturbation, they
exist in a turbulent background flow, and they propagate over a
complex topography.  It is expected that these perturbations break the
homoclinic orbit and produce chaotic trajectories.  While the
transport due to an unperturbed vortex has previously been considered
[4], the impact of the chaotic trajectories has not to our
knowledge been studied.

To construct a Hamiltonian map for an isolated vortex, a ``vortex
map'', we start with the streamfunction for an unperturbed vortex.
Many quantitatively different but qualitatively similar
streamfunctions have been used to study isolated vortices
[17, 18, 19].  Here we choose a Gaussian streamfunction, used in
[17] as their standard profile, which has the advantage of
having zero net circulation and hence leaves the flow unaffected at
infinity.  We assume the vortex has amplitude $A$ and is propagating
westward (towards negative $x$).  In a frame comoving
with the vortex the streamfunction is
$$
\psi_0(x,y) = A e^{- (x^2 + y^2)} - y,
\eqno(4)
$$
where distances are measured in units such that the vortex size is
unity and time is measured in units such that the propagation speed is
unity.  The streamfunction undergoes a saddle-node bifurcation
when  $A = A_c = e^{1/2}/\sqrt{2} \approx 1.16$; below $A_c$ there
are no fixed points while above $A_c$ there is a stable fixed point at
$x=0$, $y = y_s$ and an unstable fixed point at $x=0$, $y=y_u$.  The
fixed points are solutions of $2 A y \exp({-y^2}) = 1$ and $y_u <
y_s$.   The existence of closed streamlines and stagnation points in
ocean vortices indicates that they have amplitudes above the
bifurcation value.  Figure~1 shows $\psi_0$ for $A = 2.0$.

The next step is to pick a periodic perturbation $\psi_1$ that allows
the explicit construction of a map.  This can be done if the addition of
$\psi_1$ results in a streamfunction which has separate terms acting
at different times, each of which can be analytically integrated.  One
perturbation which accomplishes this is:
$$
\psi_1(x,y,t) = A e^{-(x^2 + y^2)} \left( k \sum_{n = -\infty}^{\infty}
\delta(t - n k^+) - 1\right), \eqno(5)
$$
where $\delta$ denotes the Dirac delta function, and $n k^+$ denotes
letting the delta-function act at time $n k + \epsilon$ and taking the
limit $\epsilon \rightarrow 0$ after doing the integration.  The total
streamfunction, $\psi = \psi_0 + \psi_1$ is
$$
\psi = A k e^{-(x^2 + y^2)} \sum_{n = -\infty}^{\infty} \delta(t - n
k^+) - y, \eqno(6)
$$
which is of the form described above since the exponential part only
acts at times $t = nk^+$, and the linear part acts during the
remainder of the time.

In Cartesian coordinates the equations of motion (3)
cannot be integrated through the delta-function because the equations
are coupled and require the value of $x$ and $y$ at the time the
delta-function acts. In polar coordinates, however, the equations of
motion are
$$
\eqalign{
\dot r &= \cos \theta, \cr
 \dot \theta &= - 2 A k e^{-r^2}
\sum_{n = -\infty}^{\infty} \delta(t - n k^+)  - \sin \theta / r, \cr}
\eqno(7)
$$
and the delta-function appears only in the $\dot \theta$ equation with
an amplitude that depends only on $r$.  To integrate the
equations from $t = n k$ to $t = (n+1)k$ one first integrates in polar
coordinates from $t = nk$ past the delta-function to, say, $t = nk + 2
\epsilon$ resulting in
$$
\eqalign{
r(nk + 2 \epsilon) &= r(nk) + O(\epsilon), \cr
\theta(nk + 2 \epsilon) &= \theta(nk) - 2 A k e^{-r^2(nk)} +
O(\epsilon).\cr}
\eqno(8)
$$
Transforming back to Cartesian coordinates, integrating to
$t = (n+1)k$, and letting $\epsilon \rightarrow 0$ results in the
vortex map:
$$
\eqalign{
x_{n+1} &= r_n \cos(\theta_n - 2 A k e^{-r_n^2}) + k,\cr
y_{n+1} &= r_n \sin(\theta_n - 2 A k e^{-r_n^2}),\cr}
\eqno(9)
$$
where the subscript $n$ refers to quantities at $t = nk$.

The streamfunction $\psi_1$ is, in the limit $k \rightarrow 0$, a
small-amplitude, high-frequency perturbation to $\psi_0$.
In addition, $\psi_1$ is a perturbation in the sense that for
small $k$ the map (9) is a finite difference approximation
to the equations of motion given by $\psi_0$ alone.  Both the standard
map [10, 11] and the traveling wave map [5] are
obtained by adding the same type of perturbation.  Note that unlike
the standard map and the traveling wave map, the vortex map is neither
periodic in $x$ nor $y$.

For very small $k$ the vortex map, like other chaotic maps, has very
thin chaotic layers which grow as $k$ increases.  The chaotic layer
which has the most significant impact on transport is the one that
appears around the stable and unstable manifolds of the unstable fixed
point. This chaotic separatrix layer is due to the breaking by the
perturbation of the  homoclinic orbit in $\psi_0$, resulting in
intersections of lobes of the manifolds creating a homoclinic tangle
[11, 20].  Figure~2 shows iterations of the vortex map
(9) for initial conditions in the separatrix layer.
The sharp boundary of the chaotic region due to KAM curves and the
existence of islands and lobes is clearly seen.  The importance of the
separatrix layer is that it allows the vortex to trap and release
fluid parcels, a phenomenon which is absent in $\psi_0$.  The trapping
manifests itself by parcels rotating about the core of the vortex for
a while before being released.  Since the vortex is propagating, this
trapping results in spatial transport of fluid parcels.

The transport in regions outside the chaotic layer is essentially
unchanged from the unperturbed case.  Inside the vortex there remains
a core of fluid where particles are permanently trapped, while far
from vortex particles flow past the vortex without ever being trapped.
These regions both have chaotic layers, but they do not affect the
trapping behavior of the vortex.  The core of permanently trapped
fluid is similar to that seen by Babiano, et.al. [13], in more
complex flows.  The size of the core decreases as the separatrix layer
grows.

In what follows, we shall study the dependence of transport in the
separatrix layer on the the amplitude of the perturbation $k$ and
amplitude of the vortex $A$.  As we are concerned with the effects of
small perturbations on the vortex, we shall only investigate $k$ up to
0.7.

The trapping time $\tau(y; x_0, x_1)$ is defined as the time $\tau =
n_t k$, where $n_t$ is the number of iterations for a parcel starting
at $(x_0,y)$ to reach $x_1$.  If $x_0$ and $x_1$ are on either side of
the vortex then $\tau$ is a good measure of the time spent being
carried by the vortex.  Figure~3 shows one example of $\tau(y)$.
One sees multiple peaks where parcels are trapped for very long times,
separated by regions where parcels travel relatively quickly through
the vortex.  The peaks are distributed in a self-similar manner
typical of structure in chaotic regions.

Using $\tau(y)$ and defining a threshold $\tau_0$ allows measurement
of the width of the chaotic layer: $w = y_1 - y_0$, where $y_1$ and
$y_0$ are, respectively, the largest and smallest $y$ with $\tau(y) >
\tau_0$.  We find that the measured width is relatively insensitive to
$\tau_0$ for $\tau_0$ sufficiently large.  Figure~4 shows $w(k)$
for three values of $A$, indicating that the width grows with both $k$
and $A$.  At smaller values of $k$, the growth of $w$ is extremely fast.

As a vortex travels through the fluid it picks up parcels, carries
them a while, and deposits them elsewhere.  The final spatial
distribution in $x$ of an initial small region of fluid is given by
the distribution of trapping times $p(\tau)$, with the speed of the
vortex providing the transformation between time and distance.
Figure~5 shows the distribution of trapping times for $A = 2.0$
and a range of $k$.  The distributions show a significant amount of
structure and appear to be composed of many individual steep
exponentials which merge together at larger $\tau$ resulting in a
shallow exponential tail.  This merging occurs at smaller $\tau$ for
larger $k$.  The separate exponentials all appear to have the same
slopes, and the slope of the tail appears independent of $k$.

Figure~6 shows $p(\tau)$ for a range of $A$ at fixed $k$.  One
sees that the slopes of both the individual exponentials and the
smooth tails are roughly independent of $A$.  At lower $A$ the
individual exponentials remain well resolved out to larger $\tau$.

A flow with a single unstable periodic orbit, such as $\psi_0$ with
its unstable fixed point, results in a $p(\tau)$ with a single
exponential.  The slope of that exponential for $\psi_0$ matches that
of the steep exponentials in Figures~5 and 6.  We conjecture that the
individual steep exponentials can each be associated with the slowing
down due to a single periodic orbit in the chaotic layer.  Since the
Smale horseshoe producing the chaotic layer has an infinite number of
periodic orbits with orbits of all periods [21] one expects the
fine structure of $p(\tau)$ to be quite complex.  The connection
between periodic orbits and transport is currently an active area of
investigation, see for example [22, 23].  The constancy of
the slopes in Figures~5 and 6 indicates that in some sense the
structure of the periodic orbits is independent of both $A$ and $k$.
This raises the possibility that not only is the qualitative
phenomenon of trapping independent of the details of the vortex and
perturbation, but that some quantitative aspects such as the slopes in
$p(\tau)$ are also robust.

\bigskip

\noindent{\bf IV.  Discussion}

Understanding the transport properties of structured flows is
extremely important to many geophysical problems.  The method
described here of constructing Hamiltonian maps should provide a
useful tool for the first step in reaching such understanding.
However the method is only a first step in that the map is a
significant simplification of the true flow.  First, the resulting
maps are kinematic, in that they use an assumed Eulerian evolution,
rather than the true dynamic evolution of the flow.  Indeed, the
simple flows which allow the explicit construction of maps are
typically not solutions of the Eulerian dynamical equations.  The
validity of the approach rests on the ansatz that the chosen flow
captures those aspects which are important to transport: the topology
of the streamlines in the time-independent idealized structure, and
the existence of small perturbations.  Second, the maps are
two-dimensional, and thus exclude any structures where the third
dimension is important for transport.  The degree to which transport
under the true Eulerian evolution in both two and three-dimensional
flows differs from that in simple Hamiltonian maps is an important
question which must be investigated.

There is evidence, however, that Hamiltonian maps do capture some
important aspects of transport.  del-Castillo-Negrete and Morrison
[6] used a Hamiltonian map to reproduce the transport seen
in laboratory experiments on Rossby waves in rotating fluids
[24].  Laboratory experiments of vortices on a $\beta$-plane
[18] show similar behavior to the above vortex map
(9).  In particular, the lobes seen in their experiment
match the lobes found in homoclinic tangles and visible in the vortex
map (Figure~2).  However, the lobes in the experiment contain
potential vorticity, an active rather than passive tracer, and the
lobes subsequently roll up into secondary vortices.  This roll-up is
absent from the vortex map due to the purely kinematic aspect of the
formulation.  Thus, while kinematic Hamiltonian mappings are only an
approximation to the true behavior of fluid structures, they are
relatively simple to construct for a variety of structures and allow
one to efficiently calculate transport properties.

\bigskip

\noindent{\bf V. Acknowledgements}

The National Center for Atmospheric Research is sponsored by the
National Science Foundation.

\vfil \eject

\noindent{\bf VI. References}

\noindent 1.  A.S. Bower and T. Rossby, J. Phys. Oceanogr. {\bf 19}
(1989) 1177.

\smallskip

\noindent 2.  Michael E. McIntyre, in {\it Dynamics, Transport, and
Photochemistry in the Middle Atmosphere of the Southern Hemisphere},
A. O'Neill, ed., p.1 (Kluwer Academic Publishers, Dordrecht, 1990).

\smallskip

\noindent 3.  P.L. Richardson, in {\it Eddies in Marine Science}, A.R.
Robinson, ed., p.19, (Springer-Verlag, Heidelberg, 1983).

\smallskip

\noindent 4.  William K. Dewar and Glenn R. Flierl, Dyn. Atmos. Oceans
{\bf 9} (1985) 215.

\smallskip

\noindent 5.  Jeffrey B. Weiss, Phys. Fluids A {\bf 3} (1991) 1379.

\smallskip

\noindent 6.  Diego del-Castillo-Negrete and P.J. Morrison, Phys.
Fluids A {\bf } (1993) 948.

\smallskip

\noindent 7.  Hassan Aref, J. Fluid Mech. {\bf 143} (1984) 1.

\smallskip

\noindent 8.  J.M. Ottino, {\it The kinematics of mixing: stretching,
chaos, and transport}, (Cambridge University Press, Cambridge, 1989).

\smallskip

\noindent 9.  E. Knobloch and J.B. Weiss, Phys. Rev. A {\bf 36} (1987)
1522.

\smallskip

\noindent 10.  B.V. Chirikov, Phys. Rep. {\bf 52} (1979) 265.

\smallskip

\noindent 11.  A.J. Lichtenberg and M.A. Lieberman, {\it Regular and
Stochastic Motion}, (Springer-Verlag, New York, 1983).

\smallskip

\noindent 12.  Darin Beigie, Anthony Leonard, and Stephen Wiggins,
Nonlinearity, {\bf 4}, 775, (1991).

\smallskip

\noindent 13.  A. Babiano, G. Boffetta, A. Provenzale, and A.
Vulpiani, preprint.

\smallskip

\noindent 14.  P.L. Richardson, D. Walsh, L. Armi, M Schr\"oder, and
J.F. Price, J. Phys. Oceanogr. {\bf 19} (1989) 371.

\smallskip

\noindent 15.  James C. McWilliams and Glenn R. Flierl, J. Phys.
Oceanogr. {\bf 9} (1979) 1155.

\smallskip

\noindent 16.  Richard P. Mied and Gloria J. Lindemann, J. Phys.
Oceanogr. {\bf 9} (1979) 1183.

\smallskip

\noindent 17.  Peter R. Gent and James C. McWilliams, Geophys. and
Astrophys. Fluid Dyn. {\bf 35} (1986) 209.

\smallskip

\noindent 18.  G.F. Carnevale, R.C. Kloosterziel, and G.J.F. van
Heijst, J. Fluid Mech. {\bf 233} (1991) 119.

\smallskip

\noindent 19.  X.J. Carton and J.C. McWilliams, in {\it
Mesoscale/Synoptic Coherent Structures in Geophysical Turbulence},
Jacques C.J. Nihoul and Bruno M. Jamart, eds., p.255 (Elsevier,
Amsterdam, 1989).

\smallskip

\noindent 20.  S. Wiggins, {\it Global Bifurcations and Chaos},
(Springer-Verlag, New York, 1988).

\smallskip

\noindent 21.  John Guckenheimer and Philip Holmes, {\it Nonlinear
Oscillations, Dynamical Systems, and Bifurcations of Vector Fields},
(Springer-Verlag, New York, 1983).

\smallskip

\noindent 22.  Bruno Eckhardt, Phys. Lett. A {\bf 172} (1993) 411.

\smallskip

\noindent 23.  Roberto Artuso, Giulio Casati, and Roberti Lombardi,
Phys. Rev. Lett. {\bf 71} (1993) 62.

\smallskip

\noindent 24.  J. Sommeria, S.D. Meyers, and H.L. Swinney, Nature {\bf
337} (1989) 58.

\vfil \eject

\noindent{\bf Figure Captions}

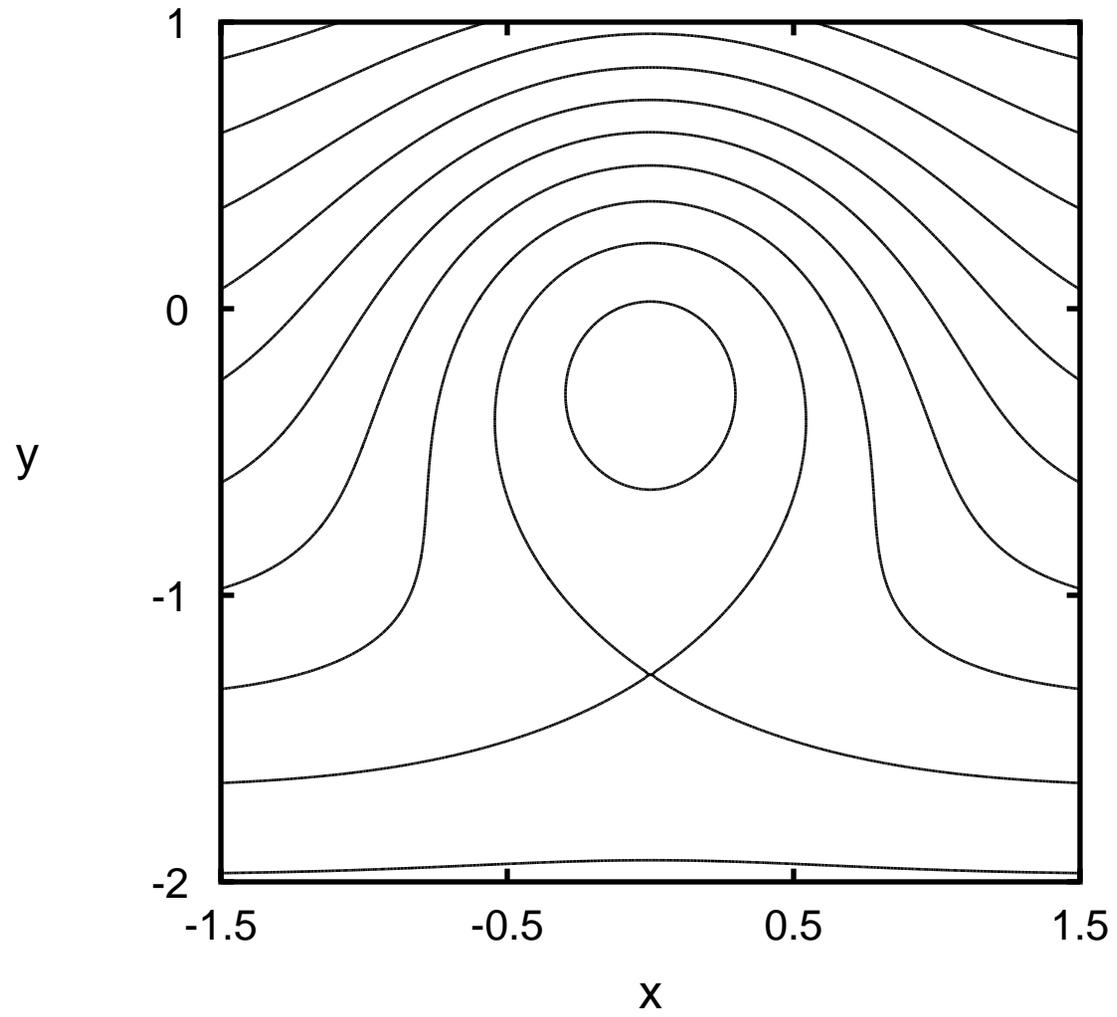
\noindent {\bf Figure 1.}  Streamfunction  $\psi_0$ with $A = 2.0$.

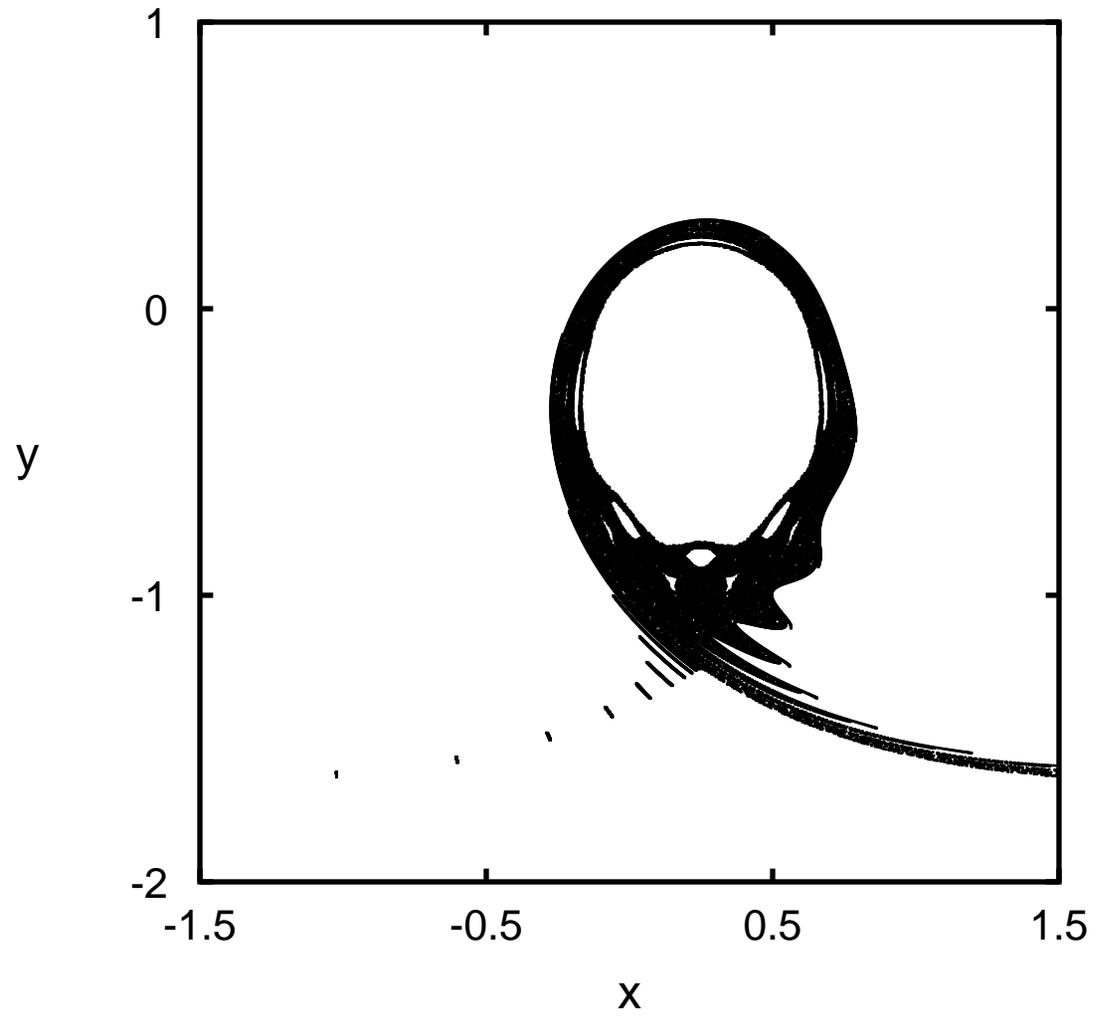
\noindent {\bf Figure 2.}  Trajectories of the vortex map with $A =
2.0$ and $k = 0.5$ for 500 initial conditions starting at $x = -1.5$
and evenly spaced across the chaotic layer.

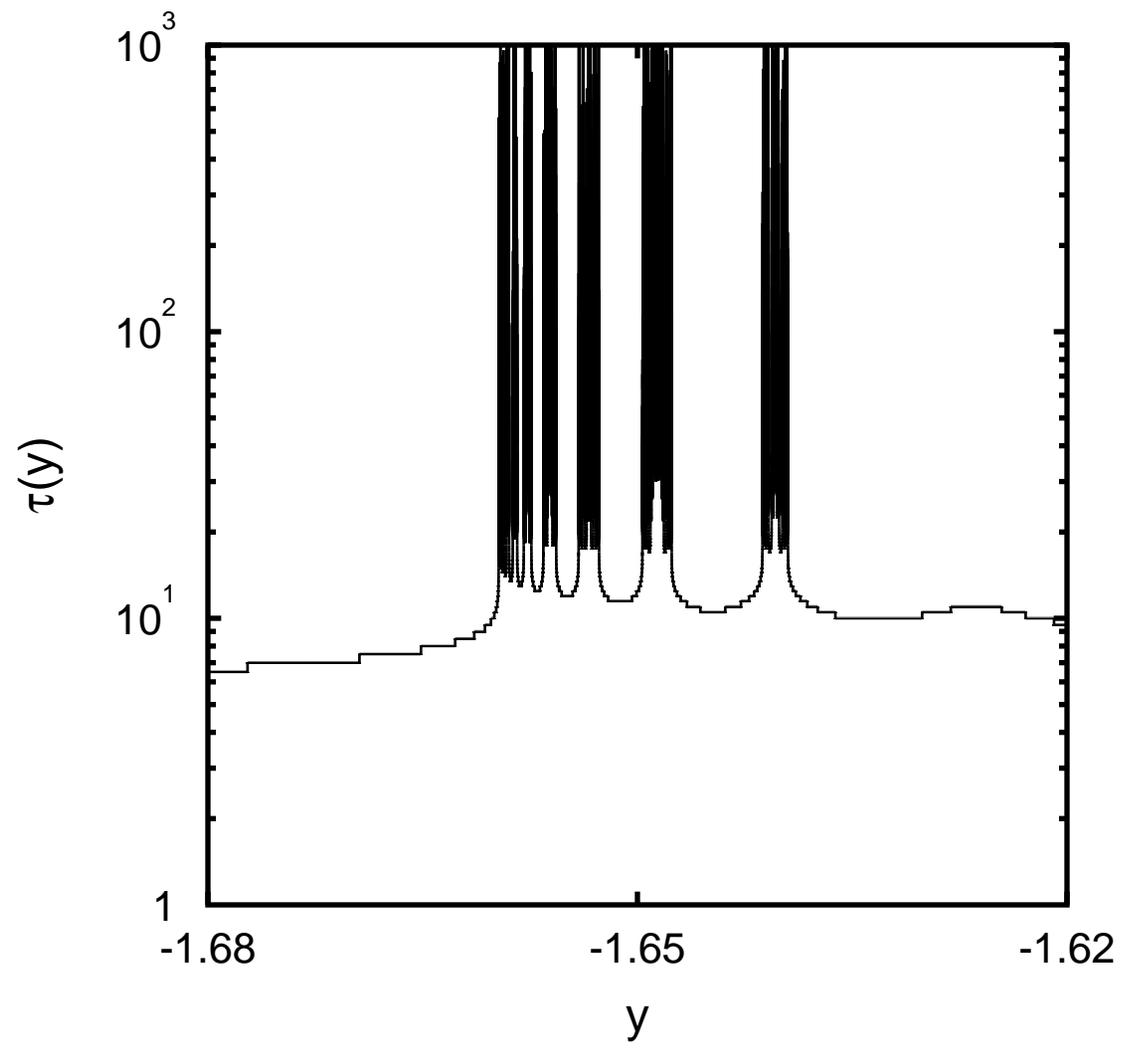
\noindent {\bf Figure 3.}  Trapping time $\tau(y; x_0, x_1)$ for $A =
2.0$, $k = 0.5$, $x_0 = -2.0$ and $x_1 = 2.0$.

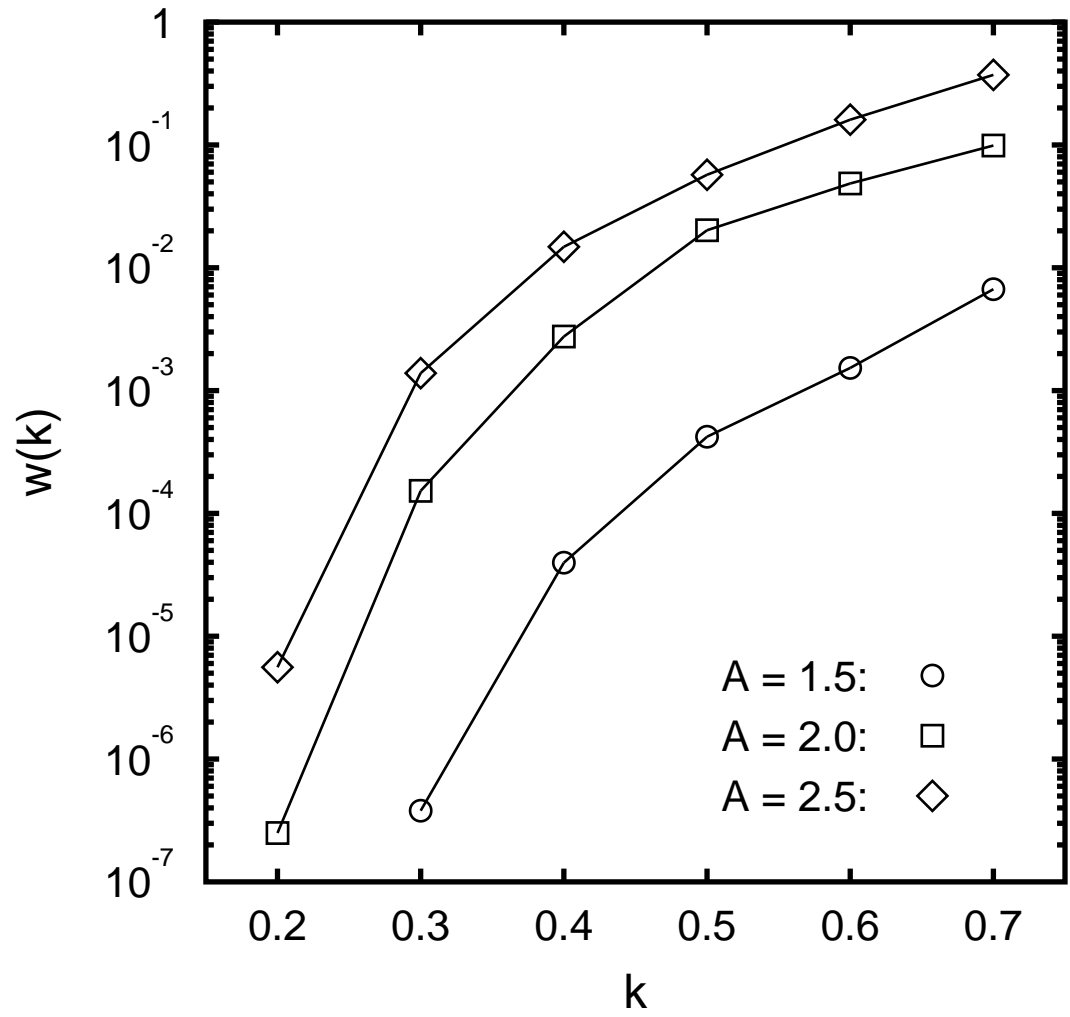
\noindent {\bf Figure 4.}  Width of the chaotic layer at $x = -2.0$
using a threshold trapping time of $\tau_0 = 50$, $x_0 = -2.0$, and
$x_1 = 2.0$.

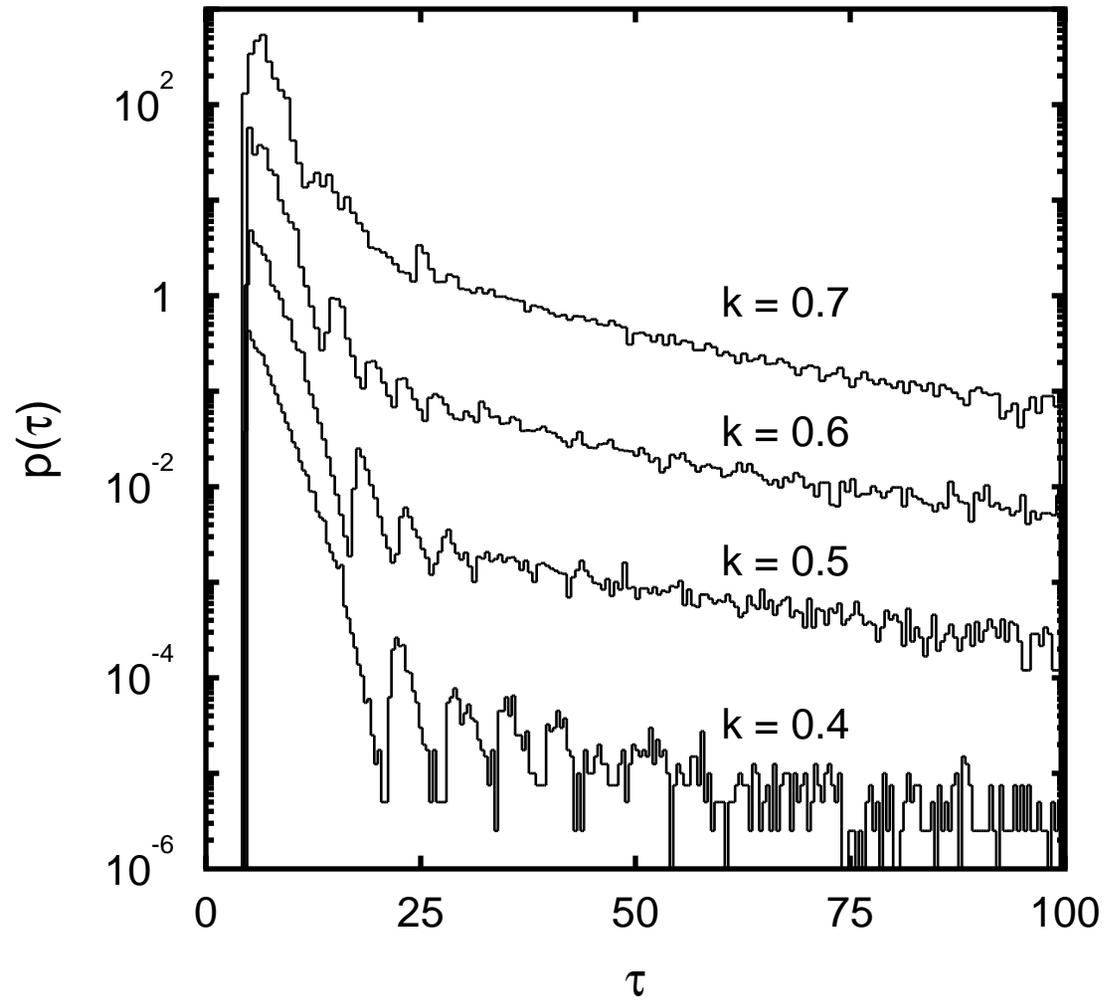
\noindent {\bf Figure 5.}  Distributions of trapping times $p(\tau)$
for $A = 2.0$, $x_0 = -2.0$, $x_1 = 2.0$ and a range $k$, each
computed from $10^6$ initial conditions evenly spaced in $y$.  The
bins in the histograms have a width of one iteration, $\Delta t = k$.
Successive distributions are shifted vertically to ease comparison.

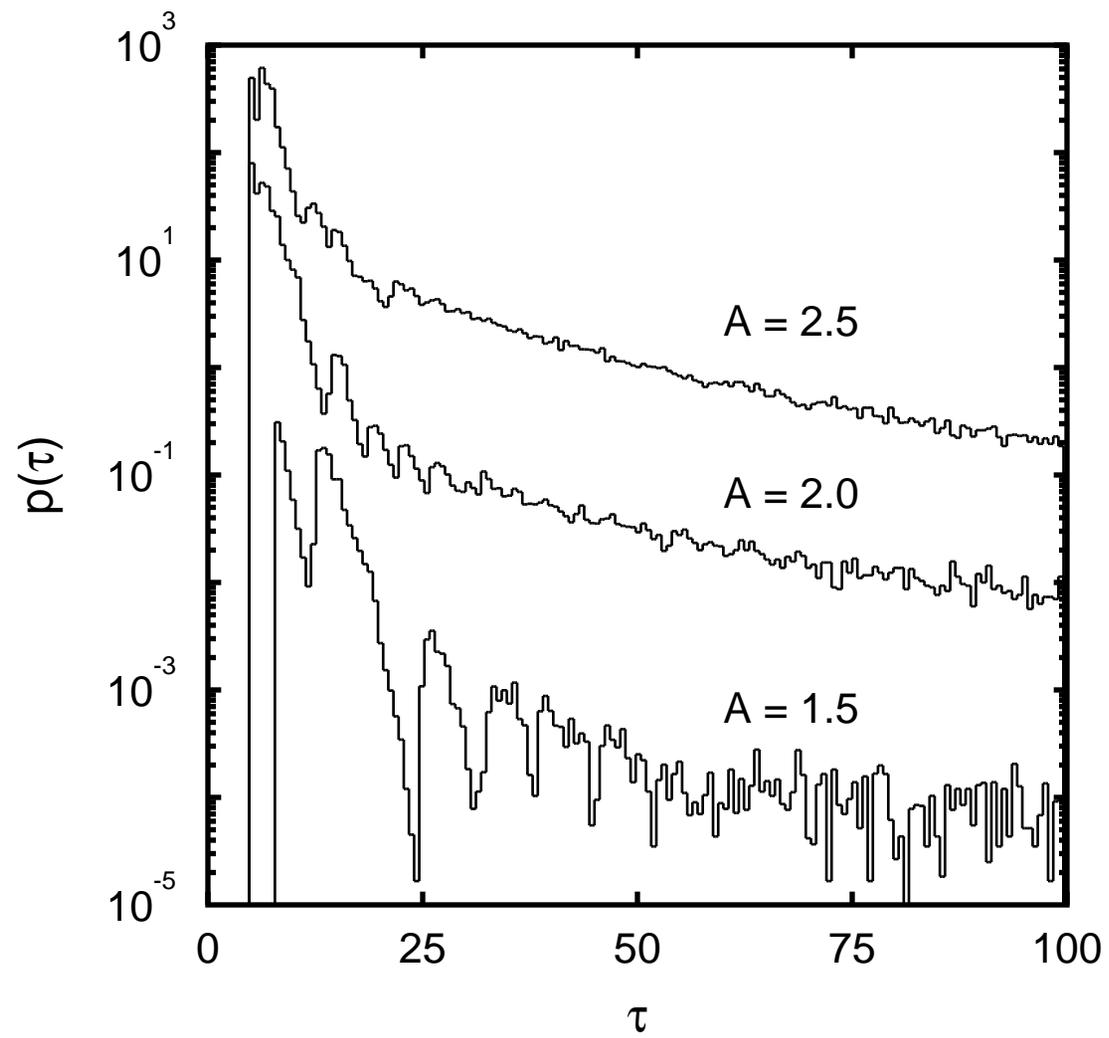
\noindent {\bf Figure 6.}  Same as Fig.5, but for $k = 0.6$ and a
range of $A$.

\end